# A Systematic Literature Review on Phishing and Anti-Phishing Techniques

Ayesha Arshad[1], Attique Ur Rehman[1], Sabeen Javaid[1], Tahir Muhammad Ali[2], Javed Anjum Sheikh[1], Muhammad Azeem[1]

[1]Department of Software Engineering, University of Sialkot, Sialkot, Pakistan
[2]Department of Computer Science, Gulf University of Science and Technology, Kuwait

Corresponding author: Attique Ur Rehman (e-mail: Attique.UrRehman@uskt.edu.pk)

*Abstract-* **Phishing is the number one threat in the world of internet. Phishing attacks are from decades and with each passing year it is becoming a major problem for internet users as attackers are coming with unique and creative ideas to breach the security. In this paper, different types of phishing and anti-phishing techniques are presented. For this purpose, the Systematic Literature Review(SLR) approach is followed to critically define the proposed research questions. At first 80 articles were extracted from different repositories. These articles were then filtered out using Tollgate Approach to find out different types of phishing and anti-phishing techniques. Research study evaluated that spear phishing, Email Spoofing, Email Manipulation and phone phishing are the most commonly used phishing techniques. On the other hand, according to the SLR, machine learning approaches have the highest accuracy of preventing and detecting phishing attacks among all other anti-phishing approaches.**
*Index Terms—* phishing techniques, anti-phishing techniques, SLR, review.

## I. INTRODUCTION

The Phishing technique and attack is a method to access sensitive and restricted information of end users by using social engineering and technology. Phishing has been declared as the number one approach used by the attackers to exploit the privacy of the internet user [1]. Most of the people who become victims, are those who do not have knowledge about phishing attacks [2]. Phishing attacks on IOT devices and machines are also growing rapidly [3]. Many security mechanisms are followed to minimize this problem but attackers are always forming ground-breaking ideas to crack undisclosed information and identities using advance technologies [4]. The most common method in phishing is sending scam emails to victims [5]. These emails are sent through the accounts which are the replicas of government authorized agencies, digital banks, electronic payment sites and digital markets like flipkart. These replicas and fraud websites gains the sensitive data from the end users through many ways [6]. These websites send the account update links, account verification emails and sometimes send prize winning messages like "congratulations you have won 10,0000 rupees, click on the link below to process" to end users by using social engineering techniques to deceive the internet users. They make them believe that those emails are coming from authorized organizations [5]. Phishing can also be done through fake phone calls for example; the person calling you present himself from any bank and ask you for your bank account details and credentials like credit card number, ATM pin code, OTP (onetime password), usernames and passwords [7]. Anti-phishing working group (APWG) reported that, 90% of the phishing outbreaks are held through HTTPS on which the data of user and browser is found. It also reported that, in the 3rd quarter of 2020, the most targeted sector is web email sites and Software-as-a-Service [8]. To minimize the phishing effects and its consequences on the users, everyone should be aware of the phishing techniques [9]. The comprehensive analysis of phishing attacks and techniques can help security developers and policy makers to develop better safety techniques and approaches [10]. Rest section 2 is presenting relevant work, where research methodology has been presented in section 3. Significance shown in section 4 and paper has been concluded a future work elaborated in section 5.

## II. LITERATURE REVIEW

When Kiren et al. [11] presented a review on different types of phishing attacks and detection techniques. Also they presented some mitigation techniques of phishing. The paper proposed that 100% accuracy to detect phishing can be made possible by using machine learning approach among all other anti-phishing approaches. Rana et al. [12] presented a review and comprehensive examination of the modern and state of the art phishing attack techniques to spread awareness of phishing techniques among the reader and to educate them about different types of attacks. The author proposed this paper to encourage the use of anti-phishing methods as well.

Justine et al. [13] proposed a phishing attack taxonomy based on E-mail. which is covering all the drawbacks of already existing phishing taxonomies. It is concluded that the proposed taxonomy has broader classification and depending on classes which are two times greater in numbers as compared to the classes of the existing taxonomies. Simono et al. [14] have briefly discussed the weak security mechanisms in password managers in android phones and these weaknesses are the cause of phishing attacks. The paper declared that there are number of issues in designs of password managers which become the cause of these attacks.

G. Jaspher et al. [15] presented different phishing attacks with latest prevention approaches in his paper. The paper shows machine learning methods to detect and distinguish phishing attacks.

Kanju Merlin et al. [16] performed the survey method to detect phishing techniques and algorithms. The survey resulted in providing many solutions and approaches to attacks detection. They showed that many of the proposed approaches are not capable enough to provide the solutions of attacks.

Aleroud et al. [17] presented a taxonomy that contains different phishing techniques its vectors and countermeasures. The paper highlighted the environments which are the most targeted ones. The taxonomy was proposed to provide the directions to design various effective anti-phishing techniques. Moreover, the proposed taxonomy was also helpful for the developers and practitioners to find out different methods and tools to combat phishing attacks.

Moul, Katelin A et al. [18] Highlighted the steps and efforts in the fight against phishing. A team consisting of some members, did these efforts and conducted different awareness sessions and workshops for the internet users to encourage the use of anti-phishing techniques and make them aware of using everything on the internet. They concluded that to combat phishing attacks, awareness sessions should be ongoing once.

Cui, Qian, et al [19] They have purposed a method of counting HTML tags used in DOM of attacks. In this method they used the concept of clustering and made clusters of attacks happening in a specific range of distance, they proposed that, these clusters can be combined and made publicly open. Their results showed that, a very large amount of newly phishing attacks can be caught by this method.

According to CYREN, in 2015 [20] there was 51% rise in phishing sites, which was threatening.

According to X-force IRIS [21]. 29 percent of attacks breach the privacy via phishing emails.45 percent of the attacks were held on business and their Emails.

Yunjia Wang et al. [22] presented a phishing prevention technique in their paper. They proposed a scheme to implement optical character recognition system on an android mobile platform. They performed experiments under hijacking attacks to check the accuracy of the proposed prevention technique. They claimed that their proposed OCR techniques are good enough to identify the phishing websites also it can overcome the problems and limitations on existing solutions.

Dr.M. Nazreen et al. [23] explained briefly the different techniques of phishing attacks. They highlighted that due to some inexperienced internet users phishing is still successful. The paper provided various phishing types used by attackers to breach the privacy.

Tanvi Churi et al. [24] presented a prototype to detect if any site is phishing site or not. They claimed in their paper that existing phishing prevention frameworks does not give 100% accuracy also existing systems are not capable of identifying phishing sites. They proposed a system which works on visual cryptography and code generation techniques which only authenticated person can breach. The proposed system generates an image which is then further divided into two shares by the visual cryptography technique then these both shares of image is combined to form an image captcha the image captcha is displayed and user is asked to match the site with image captcha to differentiate the site from phishing sites. In next step, four-digit code is generated and also authenticated by the authorized person. The method is helpful to identify the phishing sites and to protect the credentials from the unsuspected causalities.

Kang Ling Chiew et al. [25] presented the comprehensive and technical current and past phishing approaches in their paper. They claimed to provide better knowledge about the types, nature and characteristics of current and past phishing approaches through their research. The study revealed that due to the emerging technology and use of cloud computing and mobiles, the anti-phishing techniques are much needed specifically in these areas where technology is heavily involved. A great number of phishing attacks happen due to browser vulnerabilities and phishing websites.

Eduardo et al. [26] did a systematic literature review on how to face phishing attacks using latest machine learning approaches. The paper highlighted the three ways to alleviate phishing attacks which are common these days are: Necessity of awareness sessions, focused blacklists and machine learning (ML) approaches. The study revealed that from all the other machine learning approaches, Deep learning (DL) is the most emerging and efficient technique in machine learning.

Belal Amro [27] presented types of phishing attacks in mobile devices and different mitigation techniques and anti-phishing techniques. Also they provided important steps to protect against phishing in mobile systems. The paper highlighted that current anti-phishing techniques have some shortcomings which makes them less efficient in detecting phishing attacks.

Christina D [28] Stafford highlighted the factors and impacts of phishing attacks on human and how human become victim of these attacks. The research highlighted that human becomes the victim of phishing because of their own personality traits and habits such as narcissism, gullibility and habitual email use. The research revealed that from all the phishing techniques spear phishing is most targeted form of phishing.

Athulya et al. [29] discussed the different phishing attacks, latest phishing techniques used by the phishers and highlighted some anti-phishing approaches. The paper raises awareness about phishing attacks and strategies and urge the readers to practice the anti-phishing approaches. The paper proposed a phishing detection approach which helps to detect phishing website in an efficient way.

Dr Akashdeep et al. [30] Presented that phishers and cyber attackers continue to bring new strategies and tactics that are difficult to track. Cyber attackers which are using phishing attacks, whaling and spear phishing are difficult to detect and track. This can be avoided by increasing the amount of awareness about phishing among the internet users.

## III. RESEARCH METHODOLOGY

The systematic literature review is a protocol-based research methodology The paper is following the methodology introduced by kitchenham [33].

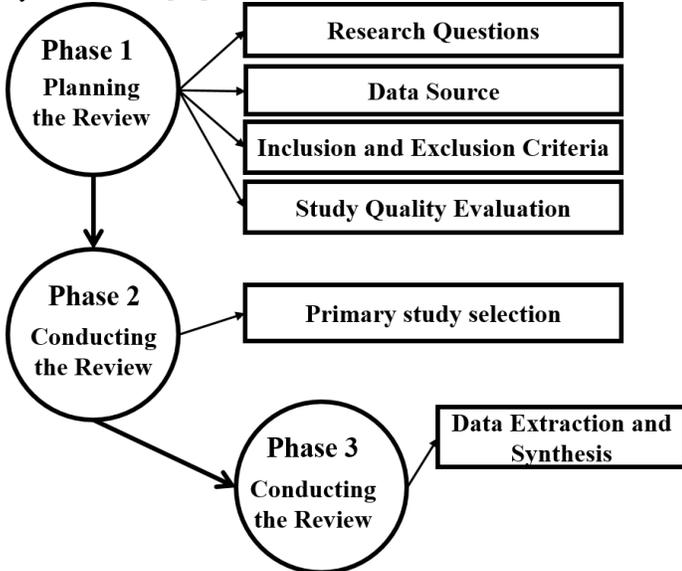

FIGURE 1: Phases of SLR

According to him the systematic literature is used for analyses, study, observe and research on a specific domain by following the inclusion and exclusion criteria [31][32][36] and is more thorough and have huge detail on a specific topic, while on the other hand informal literature review is less thorough [34]. The paper focused on the three phases of SLR i.e., planning, conducting and reporting the review.

### A. PHASE 1: PLANNING THE REVIEW

1) Research Questions: The proposed research questions for the specific topic are given below.
RQ1: What are the phishing techniques?
RQ2: What are the anti-phishing techniques?
RQ3: How effective the exciting anti-phishing techniques are?

2) Data Source: A data warehouse was selected as suggested by khan et al. [35] Table I shows the list of data warehouse.

TABLE I
DATA SEARCH STRINGS

| List | Sources |
|---|---|
| Electronic Database | ACM digital Library (ACM Digital Library) IEEE Explore(https://ieeexplore.ieee.org/) Springer Link (link.springer.com). Google Scholar (scholar.google.com). |
| Searched item | Conference, journal and books |
| Search applied on | Do not miss articles that are relevant to our study, doesn't matter if those articles do not include search keywords |
| Language | English |
| Publication period | Year 2010-2020 |

3) Inclusion Criteria: The inclusion criteria were used to identify and extract the useful literature from the search strings [37] [39]. Research articles should be in conference, book or journal. Reports that provide information about cyber-attacks and information security attacks should be included. Paper that defines different methods and techniques used to exploit privacy. Papers that define existing phishing taxonomies and prevention methods. Papers that define how effective the existing phishing techniques are: Papers from publication period (2010-2020) Papers should be in English

4) Exclusion Criteria: This criterion is used to exclude irrelevant studies from the gathered literature [31] [38] [39]. Papers that are irrelevant to the study object. Papers that do not include phishing and anti-phishing techniques. Papers that do not highlight all the research questions.

5) Study Quality Evaluation: In data extraction phase, quality evaluation (QE) were performed on the final study articles. The proposed method given by [39] were followed to make a checklist to perform QE. The QE checklist consists of three questions, each question is assigned a score as shown in Table II and presented in Appendix A. By following Shamem et al. [39] criteria, the quality of the selected study object was measured. The purpose of the quality evaluation is to provide a prized contribution to the SLR.

TABLE II
QUALITY EVALUATION CRITERIA

| Quality Evaluation Score | Quality Evaluation Criteria |
|---|---|
| QES1 | The articles were assigned 1 score that contained answer of the check list. |
| QES2 | The articles were assigned 0.5 score that contained partial answers to the checklist. |
| QES3 | The articles were assigned 0 score that's not contained the answers to the checklist questions. |

### B. CONDUCTING THE REVIEW

1) Primary Study Selection: The tollgate method presented in [32] is used to refine the selected study articles. The tollgate method consists of five phases shown in Table III. Initially 80 articles were extracted from different repositories by using search strings. The inclusion and exclusion criteria were also followed. the total of 20 articles were selected by using tollgate approach [32] [40] which is the 25% of the total extracted articles as shown in Fig 2.

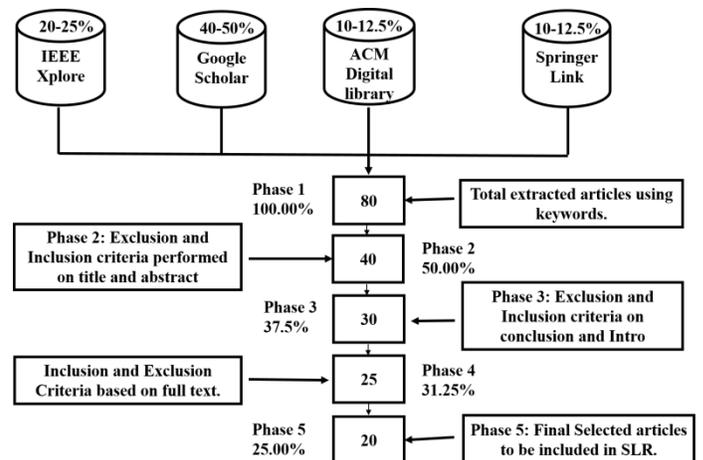

FIGURE 2: Tollgate Approach Depiction

165

## C. REPORTING THE REVIEW

1) Quality Attributes: On the basis of three research questions, the Quality Evaluation score is graded. In (Appendix A) the list of the questions with obtained score is given. The final result shows that 60% of the papers scored above 80% which means the selected articles are effective to answer the research question.

2) Temporal Distribution of the Selected Primary Studies: The publication period of selected articles is distributed in two halves. The first half consists of papers from (2018-2020) and the second half from (2013-2017). The paper from year 2018- 2020 is 75% of the total selected papers and rest of the 25% is from year 2013-2017.

TABLE III
TOLLGATE APPROACH

| Electronic Data | P1 | P2 | P3 | P4 | P5 | N=20 |
|---|---|---|---|---|---|---|
| IEEE Xplore | 20 | 10 | 8 | 7 | 5 | 25 |
| ACM Library | 10 | 5 | 3 | 2 | 2 | 10 |
| Springer | 10 | 5 | 4 | 3 | 1 | 5 |
| Google Scholar | 40 | 20 | 15 | 13 | 12 | 60 |
| Total | 80 | 40 | 30 | 25 | 20 | 100 |

## IV. RESULTS AND DISCUSSIONS

To extract the visible difference between all the phishing and anti-phishing techniques linear by linear chi-square test is used.

TABLE IV
LIST OF IDENTIFIED PHISHING TECHNIQUES

| S.NO | Identified phishing techniques | Frequency (N=20) | Percentage% |
|---|---|---|---|
| Tech1 | Email spoofing/ Email manipulation | 12 | 60% |
| Tech2 | Malvertising | 4 | 20% |
| Tech3 | Browser Vulnerabilities | 7 | 35% |
| Tech4 | Clickjacking | 4 | 20% |
| Tech5 | Cross-site Scripting(XSS)Attack | 5 | 25% |
| Tech6 | Spear Phishing | 11 | 60% |
| Tech7 | Man-in-the-Middle Attack | 4 | 20% |
| Tech8 | Phone phishing | 10 | 50% |
| Tech9 | Whaling | 4 | 20% |
| Tech10 | Pharming | 3 | 15% |
| Tech11 | Drive by Download | 2 | 15% |
| Tech12 | Vishing | 5 | 25% |

TABLE V
LIST OF IDENTIFIED ANTI-PHISHING TECHNIQUES

| S.NO | Identified Anti-phishing techniques | Frequency (N=20) | Percentage% |
|---|---|---|---|
| Tech1 | Content Filtering | 6 | 30% |
| Tech2 | OCR method in mobile | 2 | 10% |
| Tech3 | Visual cryptography and code generation technique. | 1 | 5% |
| Tech4 | Multi Factor Authentication | 10 | 50% |
| Tech5 | Machine Learning Approach | 12 | 60% |
| Tech6 | Black listing | 3 | 15% |

## A. THE LIST OF IDENTIFIED PHISHING TECHNIQUES

From total of 20 articles discussed in SLR, 12 Phishing techniques have been identified which are the answer or RQ1 which is presented in Table 5.

Tech1(Email Spoofing/Email manipulation, 60%) was identified a phishing technique in the literature. Dr M Nazreen [23] reported that email spoofing is one of the effective techniques in phishing where the sender pretends to be someone from higher authorities or from known organization to earn the trust of the victims. Justinas Rastenis et al. [13] also stated that E-mail based phishing attack is the most common one in phishing.

Tech2(Malvertising, 20%) is one of the phishing techniques in which a software is installed in the victim's system to get the sensitive information. [23]. Kang lang cheiw et al. [25] reported that, the attackers try to send advertisements to the victims. By clicking on those advertisements viruses and computer worms gets into your system to breach the sensitive data.

Tech3(Browser Vulnerabilities, 35%) was reported as a phishing technique. Kang et al. [25] underlined that the phishers are always in intent to find out the system and browser vulnerabilities to crack the systems. Chances of vulnerability in a system gets higher when a new feature or module is added to it. Merlin et al. [16] underlined that phishers try to point out illegal websites as legal by using browser vulnerabilities.

Tech4(Clickjacking, 20%) was reported as a phishing technique. Rana [12] reported that in this technique, user is forced to click on malicious links by means of closing them or giving them any kind of greed which leads to the data breaching.

Tech5(Cross-site Scripting (XSS) Attack, 25%) was identified as a phishing attack. It is used by phishers to inject malicious data into the website it happens due to the poorly developed systems [25].

Tech6(Spear Phishing, 60%) was reported as a phishing technique. In which a specific group of community is targeted like an organization or business [23] the attacker sends emails to the employees of the organization pretending themselves as some higher authority of business owner [18]. It makes victims to reply to the scam emails more intentionally.

Tech7(Man-in-the-Middle Attack, 20%) was identified as phishing attack in which a malicious user stands between the service provider and the using party which then steals the data, communication and victim's accounts credentials [12]

Tech8(phone phishing, 50%) was reported as a phishing technical approach. In which malicious attacker makes a social engineering attack by using a telephone or a mobile phone to access the delicate and confidential data from the victim either by calling, messaging or sending advertisements and malicious links to the victim [12].

Tech9(whaling, 20%) G. jaspher [15] reported that whaling is a phishing attack in which an attacker acts as a senior member of any organization and targets the other employees of the company or organization to steal the sensitive information.

Tech10(pharming, 15%) was stated as a phishing technique from the literature in which a code or link is sent at the victim's email address which can modify the system's localhost data. This makes system or website to redirect to the malicious sites [29]. Tech11(Drive by download, 15%) Merlin [16] reported that drive by download is a phishing technique that inserts malicious codes and viruses into a system by using vishing technique. Tech12(Vishing, 25%) Ketaline [18] reported that it is kind of voice phishing in which attackers use voice to manipulate data and ask the user to do malicious acts. For example, you receive a voicemail "Your Microsoft window license key has been expired please call 866-978-7540. It's a scam voicemail and asks you for your credentials upon calling.

### B. THE LIST OF IDENTIFIED ANTI-PHISHING TECHNIQUES
Six Anti-phishing techniques are identified which is given in Table VI which are the answer of RQ2 and RQ3.
A-Tech1(Content-filtering, 30%) has been reported as an anti-phishing technique in which the emails get filtered using machine learning techniques to discriminate them from scam emails [23].
A-Tech2(OCR method in mobiles, 10%) was identified as an effective anti-phishing technique. Wang et al. [22] reported that the technique has four phases, first it makes a test server to test the URLs. In the first phase it connects the URL to the mitmproxy. In the second phase it interrupts the traffic and send it to the test server. Finally, it executes the results in the mitmproxy by implementing OCR in the test server.
A-Tech3 (Visual Cryptography and code generation Techniques, 5%) was identified as one of the anti-phishing technique by Tanvi [24] which reported that it is an effective prevention technique to keep the internet users far away from the phishers.
A-Tech4 (Multi Factor Authentication, 50%) was identified as an anti-phishing techniques. Ketaline [14] represented the Multi Factor Authentication technique uses two or more authentications to login into the accounts/systems. One is password and other is code generated by an app through SMS, phone calls or emails. By this method only authenticated person can login into his account.
Tech5(Machine Learning Approach, 60%) was identified as one of the most effective anti-phishing technique.
Eduardo et al. [26] presented a deep learning approach using Machine Learning. The technique was used to observe the activities in operating system. If any suspicious act or software have been detected using the rules of machine learning, then it passed on to the deep learning for further evaluation phases.The most accurate method in deep learning is LTSM with 98% of accuracy. Merlin et al. [16] concluded that not all machine learning approaches are effective anti-phishing techniques. Rana et al. [12] reported that if enough training is provided, the machine learning approaches can even detect zero-day phishing attacks. G. Jaspher et al. [15] reported that the machine learning approach have accuracy of 98.4% and is the highest one from all other approaches.
A-Tech6 (Black listing, 15%) was reported as an anti-phishing technique. A group of people or community gather all the phishing sites in a single platform which is then provided to the clients and users [23].

### C. CRITICALLY IDENTIFIED PHISHING AND ANTI-PHISHING TECHNIQUES
To identify the critical, most effective and used phishing techniques, a criterion is followed which stated that, a phishing technique is classified as important if it is mentioned in the SLR with the percentage of 50% or greater. The criteria have been followed by some of the researchers [35].

### D. CRITICAL PHISHING TECHNIQUES
By following the above criteria, it has been noticed that Tech1(Email Spoofing/Email manipulation, 60%), Tech6(Spear Phishing, 60%) and Tech8(phone phishing, 50%) have been identified as the most used and effective phishing techniques according to the total selected articles for the systematic literature review.

### E. CRITICALL ANTI-PHISIHING TECHNIQUES
By following the above criteria, A-Tech4 (Multi Factor Authentication, 50%) and (Machine Learning Approaches, 60% are the most commonly used anti-phishing techniques according to the literature.

## IV. SIGNIFICANCE
The study hopes to inspire the security managers and policy makers to make better security systems and algorithms to combat phishing using state of the art anti-phishing techniques.

## V. CONCLUSION AND FUTURE WORK
This study is based on Systematic Literature Review (SLR) on different types of phishing and anti-phishing techniques. The study is proposed to provide healthier understandings among the readers, internet users and the security managers about the phishing and anti-phishing techniques. For this purpose, phishing and anti-phishing techniques have been examined, analyzed and surveyed. On the basis of the SLR it has been evaluated that most used phishing techniques are spear phishing, email based attacks and phone phishing while the effective and used anti-phishing techniques are machine learning. Deep learning in machine learning is playing a vital role in it. The main aim of this study is to provide a holistic understandings and knowledge about current and most used phishing and anti-phishing techniques.
In future, it is expected to research and predict which anti-phishing technique would be useful and effective to combat which type of phishing attack. For example, to combat spear phishing, which anti-phishing technique would be most effective one. Prediction would be based on data sets.

TABLE VI
APPENDIX A

| ID | Ref | QE | QE2 | QE3 | Total Score | (N=3) |
|---|---|---|---|---|---|---|
| RP1 | [22] | 1 | 1 | 1 | 3 | 100% |
| RP2 | [23] | 0.5 | 0 | 1 | 1.5 | 50% |
| RP3 | [24] | 1 | 1 | 1 | 3 | 100% |
| RP4 | [25] | 1 | 1 | 1 | 3 | 100% |
| RP5 | [26] | 1 | 0.5 | 1 | 2.5 | 83% |
| RP6 | [27] | 1 | 0.5 | 1 | 2.5 | 83% |
| RP7 | [28] | 0.5 | 1 | 0.5 | 2 | 66% |
| RP8 | [29] | 1 | 0.5 | 2 | 2.5 | 83% |
| RP9 | [30] | 1 | 1 | 1 | 3 | 100% |
| RP10 | [18] | 1 | 0.5 | 0.5 | 2 | 66% |
| RP11 | [20] | 1 | 0.5 | 0.5 | 2 | 66% |
| RP12 | [11] | 1 | 1 | 1 | 3 | 100% |
| RP13 | [13] | 1 | 1 | 1 | 3 | 100% |
| RP14 | [16] | 1 | 0.5 | 1 | 2.5 | 83% |
| RP15 | [21] | 1 | 0.5 | 0.5 | 2 | 66% |
| RP16 | [14] | 1 | 0.5 | 0.5 | 2 | 66% |
| RP17 | [12] | 1 | 0.5 | 0.5 | 2 | 66% |
| RP18 | [17] | 1 | 1 | 1 | 3 | 100% |
| RP19 | [19] | 1 | 0.5 | 0.5 | 2 | 66% |
| RP20 | [15] | 1 | 1 | 1 | 3 | 100% |

## ACKNOWLEDGMENT

This work was supported by Kuwait Foundation for the Advancement of Science (KFAS) under Grant PR17-18-IQ-01.